\documentclass{agujournal2019}

\usepackage{url} 
\usepackage{soul}
\usepackage{lmodern}
\usepackage{tabularx}

\usepackage{siunitx}[=2021-04-09]    
\usepackage{graphicx}
\usepackage{amsmath}
\usepackage{booktabs}
\usepackage{multirow}
\usepackage{xspace} 
\newcommand{\geoclaw}{GeoClaw\xspace}
\newcommand{\hecras}{HEC-RAS\xspace}
\newcommand{\clawpack}{Clawpack\xspace}

\newcommand{\Fig}[1]{Figure \ref{fig:#1}}

\newcommand{\Tab}[1]{Table \ref{tab:#1}}
\newcommand{\Sec}[1]{Section \ref{sec:#1}}

\newcommand{\Sfx}{\ensuremath{S^f_x}}
\newcommand{\Sfy}{\ensuremath{S^f_y}}

\newcommand{\manningsn}{Manning's $n$\xspace}

\newcommand{\ignore}[1]{}

\journalname{}

\begin{document}

\title{Simulating the 1976 Teton Dam Failure using \geoclaw and \hecras and comparing with Historical Observations}

\authors{H. R. Spero\affil{1}, D. Calhoun\affil{2}, and M. Schubert\affil{3}}

\affiliation{1}{Boise State University, Department of Geosciences}
\affiliation{2}{Boise State University, Department of Mathematics}
\affiliation{3}{HDR Engineering Inc., Boise, Idaho}

\correspondingauthor{Hannah Spero; Current Affiliation: University of Notre Dame}{Email: hspero@nd.edu}

\section*{ORCiD Identifiers}
\begin{itemize}
    \item H. R. Spero: 0000-0002-0912-9340
    \item D. Calhoun: 0000-0002-6005-4575
    \item M. Schubert: 0000-0001-5509-7637
\end{itemize}

\begin{keypoints}

\item GeoClaw results within acceptable range for dam failure downstream modeling with comparison to historical data and HEC-RAS results.
\item Use of Lagrangian gauges to track downstream eddying in GeoClaw software.
\item Teton Dam failure benchmark problem used for comparison of software and prepared for future dam failure modeling applications.
\end{keypoints}

\section*{Keywords}
\geoclaw; numerical flood modeling; Teton Dam; \hecras; dam failure



\begin{abstract}

Dam failures occur worldwide, often from factors including aging structures, extreme hydrologic loading, and design oversights related to the changing climate. Understanding and mitigating risk to downstream inhabited areas require developing and improving low-cost high-fidelity tools, such as numerical models, which allow emergency managers to predict the consequences of dam failures better. Two-dimensional (2D) depth-averaged hydraulic models can provide valuable insights into the importance of breach parameters or downstream flow characteristics, but historical studies considering historic failures using real topographies are less common in literature. This study compares \geoclaw, a 2D hydraulic model with adaptive mesh refinement capabilities, to an industry-standard software \hecras (Hydrologic Engineering Center - River Analysis System) using the 1976 Teton Dam failure as a case study. The suitability of \geoclaw for dam failure modeling is determined based on its capability to resolve inundation extent and flood wave arrival times. This study performs sensitivity analyses of the \hecras model to compare an instantaneous dam breach assumption with a time-dependent breach formation for quantifying the model uncertainty. We find the 2D \geoclaw dam-break model results compare reasonably with historical gauge records and field observational data and \hecras results. The model demonstrates stability and relatively low computational costs. Our findings highlight opportunities for future work, with the \geoclaw software performance supporting continued studies to evaluate performance. Outcomes of this study will assist dam owners, floodplain managers, and emergency managers by providing an additional tool for estimating the impacts of dam failures to protect lives and infrastructure downstream. 

\end{abstract}

\section{Introduction}

\sisetup
{   
    group-separator = {,}
}

Dam failures have caused some of the most significant disasters associated with the failure of human-made systems, and the aging dam population is expected to increase by 65\% in failures in the next decade \cite{asdso_2021}. Although dam failures are relatively rare, with one large-scale dam failure per year, climate change studies suggest catastrophic dam failures are likely to become more common due to increased hydrologic loading, as most of the aging dam population was not designed to consider global warming; earthen dams being especially vulnerable \cite{Stanford_2018,loza2021literature,boulange2021role}. The United States (US) has \SI{91000} dams that impound \SI{965606}{\km} of rivers, with an average of ten reported dam failures occurring every year \cite{asdso_2021}. Despite advancements in technology and current monitoring tools in place, catastrophic dam failures continue to occur. Two recent examples are the Edenville and Sanford Dam failures in Michigan, United States \cite{embankments2021edenville}.

Computational models are crucial tools for assessing the downstream consequences of potential catastrophic dam failures. While a fully three-dimensional (3D) simulation of a flooding event would be prohibitive for all but the fastest supercomputers, two-dimensional (2D), depth-averaged numerical models have been shown to provide robust results that agree well with observations and can be run efficiently on standard laptops or desktop workstations \cite{neelz2010benchmarking}. Because the downstream flow conditions are often physically complex however, it is critical to investigate the importance and sensitivity of key parameters such as roughness coefficient and dam failure mode in these 2D hydraulic models. Software validation and documentation of model sensitivity provides the end user confidence in model results \cite{neelz2010benchmarking, psomiadis2021potential}. 

Many studies have generally assumed that an instantaneous dam breach can be used instead of a more realistic time-dependent dam breach \cite{wurbs1987dam, valiani2002case}. However, ongoing research is investigating the validity of this assumption due to limited scientific evidence to substantiate that assumption with earthen dam breaches \cite{yi2011dam, bhandari2017one}. We test the validity of the instantaneous breach assumption by comparing the downstream consequences between a time-dependent and an instantaneous dam failure mode for the Teton Dam failure. 

This study compares the research code \geoclaw developed initially at the University of Washington \cite{berger2011geoclaw,george2011adaptive,turzewski2019geomorphic}, to \hecras (Hydraulic Engineering Center - River Analysis System), developed by the Army Corps of Engineers \cite{brunner2002hec,urzicua2021using}. A key advantage of the \geoclaw software is that it can adaptively refine the numerical mesh used to track flood progression over large topographic domains. \hecras conversely has a sophisticated Graphical User Interface (GUI) and advanced tools for implementing different modes of dam failure and roughness coefficients, for example. \geoclaw has been validated for use in tsunami modeling, it has not been widely used for dam failure modeling, so this study tests the appropriateness of using \geoclaw for academic dam failure studies, in comparison to the \hecras software, which serves as an industry standard for dam failure modeling.

Actual dam failure events can offer a robust source of observational data for use in validating high-performance computational codes. Examples of commonly used dam failures events used as software benchmark problems include the Malpasset dam failure, Baldwin Hills reservoir failure, and Vajont dam failure case studies \cite{aureli2021review, begnudelli2007simulation, bosa2011shallow, urzicua2021using, hervouet1999malpasset, biscarini2016simulation, valiani2002case, GALLEGOS20091323}. 
In this study, we use the 1976 Teton Dam failure (US), an event that predominantly formed the basis for the understanding of dam failures \cite{chadwick1976report, reclamation2006,asdso_2021}. The Teton Dam domain provides complex topography to challenge 2D solvers and has a wealth of data associated with the event because of its historical importance. The Teton Dam failure is a demanding and unique case study for computational code comparison that requires simulating flood wave propagation over complex terrain. 

This study answers two guiding questions through an examination of the hydraulics and a thorough comparison of two 2D modeled downstream flow regimes of the Teton Dam failure:

\begin{itemize}
\item How does \geoclaw compare with \hecras using the Teton Dam benchmark problem? To answer this question this study uses a comparison criteria including flood wave arrival times, lateral inundation extent, and flow depths. 

\item How sensitive is the Teton Dam \hecras model and what is the model uncertainty? To answer this question four sensitivity analyses were performed evaluating the models sensitivity to (i) computational mesh cell size (ii) Manning's roughness coefficient (iii) the time-dependent or instantaneous dam breach assumption (iv) reservoir volume size. 

\end{itemize}

\section{Background}

\subsection{Teton Dam Site and Historic Failure}
The Teton Dam geometry and flood were well-documented by the Bureau of Reclamation (Reclamation), the United States Geological Survey (USGS), and local residents, allowing for a robust account describing the event \cite{reclamation2006,chadwick1976report,carter1976field,usgs1976survey}.

This event has been studied using one-dimensional models and has helped form the basis for the understanding of earthen dam failures \cite{blanton1977flood, snyder1977flood, brown1977flood, fread1977flood, thomas1977flood, macchione1990floods,gundlach1977guidelines,balloffet1982numerical}.  This study contributes new analyses for determining the numerical algorithms' effectiveness for dam-break modeling \cite{aureli2021review}. Furthermore, the data from this paper provides documentation of the consolidated Teton Dam failure data for future work in preparing the Teton Dam failure as a more common a benchmark problem for code comparison \cite{tdamrepository}.

The Teton Dam Site consists of the Teton River Canyon (\SI{21}{\km} upstream from Rexburg, Idaho) and the Teton Dam in Eastern Idaho, US.  The Teton River canyon is narrow at the upstream end, \SI{30.5}{\km} from its mainstem near Victor, ID. The canyon becomes gradually wider downstream with a decreasing slope \cite{magleby1981post, randle2000geomorphology, pierce1992track, williams1982cenozoic}. Reclamation commissioned the embankment dam to provide irrigation and flood control along with wildlife mitigation measures \cite{chadwick1976report}.  At time of failure, the Teton Dam was one of the tallest dams in the US in 1976, with a crest height of \SI{93}{\m} at elevation \SI{1626}{\m} above mean sea level (MSL). The length of the crest of the dam was \SI{914.4}{\m}, and the width of the crest was \SI{10.7}{\m}. Although the dam was designed to store a volume of \SI{3.55e8}{\cubic\m}  (\SI{288000} acre-ft, it only reached \SI{2.89e8}{\cubic\m} (\SI{234260} acre-ft) before failure \cite{reclamation2006}. At the time of failure, the Teton Reservoir had \SI{82}{\m} Water Surface Elevation (WSE) above the river bed \citeyear{reclamation2006}. 

At 7:30 on 5 June 1976, a piping failure occurred in the earthen dam when the underlying porous basaltic rock allowed water to seep through the embankment carrying away the core material under extreme pressures, resulting in a catastrophic breaching of the dam \cite{chadwick1976report}. The resulting flood extended \SI{250}{\km} downstream where the American Falls Reservoir captured the flow (6 June 1976). The flooding destroyed downstream infrastructure and caused 11 deaths \cite{nagel2021approaching, chadwick1976report}. From Reclamation analyses, the significant internal destruction of the dam occurred over the period of one hour between 11:00 - 11:57 on 5 June 1976 \cite{solava2003lessons, chadwick1976report}. 

\subsection{A Computational Model of Dam Failure}

\subsubsection{Shallow Water Equations Application in Dam Break Modeling}

It is widely accepted that the shallow water wave equations (SWE) are an appropriate mathematical formulation for resolving dam failure flood regimes, beginning with critical work in the 1970s \cite{martin1971finite,strelkoff1977comparative,keefer1977qualitative,shigeeda20012,kocaman2021experimental}. General agreement exists that the SWE can be used to describe dam-break waves over natural topography as they combine computational efficiency and accurate reconstruction of real-world flow regimes \cite{balloffet1982numerical,hervouet1999malpasset}. However, the SWE do not contain non-hydrostatic terms and therefore cannot explicitly represent extreme vertical accelerations directly at the dam-break site. Downstream, where a laminar, hydrostatic flow properties are valid, accurate flow fields can recovered and can accurately estimate flow regimes \cite{garcia2009dam}. Additionally, although specific 3D models can simulate dam-breaks (including depicting pressures using non-hydrostatic terms), the computational cost of a 3D model for simulating dam-break phenomena is substantial. For example, 3D models often require higher resolution topography, they solve additional terms at each time step, and require technological advances in high-performance computing or access to computing resources one might not have for parallel processing \cite{app8122456, capasso2021numerical}. 

The mathematical description of the SWEs are expressed in conservation form as 
\begin{equation}
\begin{aligned}
h_t + \nabla \cdot \mathbf u h & = 0 \\
(h\mathbf u)_t + \nabla \cdot \left(h \mathbf u \otimes \mathbf u  
+ \frac{1}{2} g h^2 \mathcal I\right) & = gh\nabla b + \mathbf S
\end{aligned}
\end{equation}
where $h(x,y,t)$ is the height of the water column, $\mathbf u = (u(x,y,t), v(x,y,t))$ are depth averaged horizontal velocities, and $b(x,y)$ is the topography.  Additional source terms $\mathbf S^f \equiv (\Sfx, \Sfy)$ model friction with the floodplain and are typically expressed as
\begin{equation}  
\begin{aligned}
\Sfx & = n^2 u\sqrt{u^2+v^2} h^{-\frac{4}{3}} \\
\Sfy & = n^2 v\sqrt{u^2 +v^2} h^{-\frac{4}{3}}
\end{aligned}
\label{eqn:friction}
\end{equation}
where $n$ is the \manningsn roughness coefficient. 

For dam failure modeling, several computational models implement the SWEs for simulating downstream flooding dynamics allowing for hazard assessment, mitigation planning, and academic studies of flow regime \cite{yakti20182d, shrestha2020understanding, GEORGEanila2015853, kumar2017literature}. Below, we describe the \geoclaw, and \hecras models with additional details are summarized in \Tab{software}.

\subsubsection{\hecras} 
First released in 1995 by the US Army Corp of Engineers (USACE) \cite{brunner2002hec}, \hecras v.5.0.7 \citeyear{hecras2019release} has the capability of performing 2D computations based on depth-averaged flow \cite{brunner2016benchmarking, hecras2019release}. \hecras uses an implicit finite volume method and can take larger time steps than would normally be required by an explicit method \cite{hecrasterrain}.
Recent work includes simulation of a levee breach and resulting inundation, where a combined 1D/2D \hecras approach was used and illustrated good agreement with observed data, and complex 2D models \cite{dasallas2019case}. 
A similar hydrodynamic approach was used, based on the \hecras v.5.0.7 model to reproduce the Ukai Dam (India) flood event \cite{patel2017assessment}. 
This simulation highlighted the broad capabilities of \hecras 5.0.7 for flood modeling and inundation mapping studies from dam failure \citeyear{patel2017assessment}. 
Additionally, \hecras has been used as a standard benchmark tool to test the performances of other models as it is both open-source and easily accessible through a Graphical User Interface (GUI) \cite{costabile2020performances, brunner2016benchmarking}. 

\subsubsection{\geoclaw}
The \geoclaw software v.5.8 (2020) is part of the open-source software package \clawpack (Conservation Laws Package).  \geoclaw has been used in previous studies for storm surge, outburst floods, debris flow, \cite{mandli2014adaptive,george2011adaptive,turzewski2019geomorphic,macinnes2013comparison,arcos2015validating}, and has been extensively tested and validated for tsunami simulations \cite{gonzalez2011validation}.  Based on a high-resolution finite volume algorithm described in \cite{le:2002}, \geoclaw is unique in that it can dynamically adapt mesh resolution to follow solution features of interest.  In dam failure modeling, \geoclaw can be tuned to use high-resolution grids only in flooding regions where the water column height is larger than zero. Using this refinement strategy, \geoclaw can efficiently solve multi-scale hydrodynamic flow problems on much larger domains than would typically be feasible for uniformly refined meshes.  \geoclaw has been used for at least one dam failure problem previously and shown to produce results with good agreement with observational data \cite{george2011adaptive}. Inputs for \geoclaw are user-defined topography, reservoir parameterization (water surface elevation, boundary), dam failure specifications (height, location, timing), and a roughness parameter. One of the chief contributions of this study is to provide more evidence that \geoclaw should be considered further for dam failure modeling.
   
\begin{table}
\caption{Comparison of \geoclaw software to \hecras software with a focus on components integral to dam failure numerical modeling such as numerical schemes, mesh shape, and adaptive refinement capabilities \cite{clawpack, berger2011geoclaw, arcos2015validating, brunner2002hec, brunner2016benchmarking}. Key differences accounted for in this study include the numerical schemes, interactive GUI, and mesh parameterization.}
\begin{center}
\begin{tabular}{rcc}
\toprule
Characteristic & \geoclaw & \hecras \\
\midrule
 Developer & Univ. of Washington & U.S. Army Corps Engineers \\
 Adaptive Mesh Refinement & Yes & No \\ 
 Discretization & Explicit FV & Implicit/Explicit FV \\  
 Shock-Capturing & Yes & Yes  \\
 1D/2D Linkages & No & Yes  \\
 Parallel Capabilities & Yes & Yes \\  
 Mesh Shape & Cartesian & Poly. Cells, User-Defined \\
 Interactive GUI Interface? & No & Yes \\
 Variable Manning's Roughness & Yes & Yes \\
 \bottomrule
\end{tabular}
\end{center}
\label{tab:software} 
\end{table}

\begin{table}[!b]
\caption{Historical data for downstream locations with corresponding flood depth, arrival times (5 June 1976), and distance from the dam; sourced from historical records \cite{chadwick1976report,usgs1976survey}.}
\begin{tabular}{rrrrr}
    \toprule
    Location & Distance [1] & Arrival Time [2] & Peak Flow & Max. Depth [3] \\
    \midrule
    Teton Canyon & 4 km       & 12:05       & \SI{65129}{\cubic\meter/s} & 15 m \\
    Teton Canyon Mouth & 8 km & 12:10-12:20 & X         & 12 m \\
    Wilford      & 13.5 km    & 12:45       & X         & 4 m \\
    Teton Town   & 12.9 km    & 12:30       & \SI{30016}{\cubic\meter/s}    & 3 m \\
    Sugar City   & 19.8 km    & 13:30       & X         & 3 m \\
    Rexburg      & 24.6 km    & 14:30       & X         & 2.5 m \\
    \bottomrule
\end{tabular}
\begin{tablenotes}
\footnotesize \emph{ 
(1) where 'distance' refers to the distance from the dam to the historical observation location 
(2) where 'arrival time' refers to the flood wave arrival time on 5 June 1976 in 24-HR clock time
(3) where 'max. depth' refers to the maximum flood wave depth in the historical observation location}
\end{tablenotes}
\label{tab:historicdata}
\end{table}

\subsection{Previous Work}
Researchers have developed a wide range of computational models for dam failure.  Below, we describe computational models of the Teton Dam and Malpasset dam failures. 

\subsubsection{Teton Dam Failure}

\paragraph{Bureau of Reclamation 1977}
Reclamation used a preliminary version of \hecras, called USTFLO  (Gradually Varied Unsteady Flow Profiles model) to investigate the Teton Dam failure \cite{rec_1980_land}. Reclamation used a 1D model with horizontal water surface traverse to the flow to model the failure \cite{gundlach1977guidelines}. The model was parameterized with an instantaneous flood development sequence and full dynamic routing to perform breach analyses \cite{rec_1980_land}. The 1976 Reclamation data sets (\Tab{historicdata}) can be compared with USTFLO Reclamation 1977 study results at \citeA{gundlach1977guidelines}.

\paragraph{Balloffet and Scheffler 1982 }
Balloffet and Scheffler applied a 1D finite-difference (FD) model to simulate the Teton Dam failure flood in an explicit scheme (\citeyear{balloffet1982numerical}). This study used a network of channels and reservoirs, considering the progression inundation and drying of the floodplain, to model the flood. Their results indicate that a 1D-FD model using network analysis allows for a more accurate representation of flood propagation than the methods of previous studies, which used 1D models producing averaged stage and discharge across the floodplain \cite{gundlach1977guidelines}.

\paragraph{Idaho National Laboratory 2015 }
Idaho National Laboratory (INL), Boise State University, and Neutrino Dynamics Inc. modeled a hypothetical Teton Dam failure. The study parameterized the Teton Dam breach with a Teton Reservoir WSE of \SI{115}{\meter}, compared to the historical \SI{82}{\meter} WSE \cite{randle2000geomorphology}. The scenario models a hypothetical Teton Dam Failure and the resulting inundation of a fictitious nuclear power plant. The inundation at INL was modeled using 3D Smoothed Particle Hydrodynamics (SPH) - a robust Lagrangian approach for simulating fluid flows. The breach was exaggerated to show a dramatic coupling of the flood \cite{smith2015flooding}. This coupling effort was modeled in three stages: (1) 2D \geoclaw simulation, (2) 3D domain of Neutrino Flow, and (3) Inflow to the 3D domain. The \geoclaw computational model used an instantaneous dam failure with a vertical wall of water where the dam was located at an initial water height of \SI{115}{\meter} from ground level \cite{smith2015flooding}. The \geoclaw results were one-way coupled to the 2D DualSPHysics SPH open-source code coupled with the 3D SPH area \cite{smith2015flooding}. This work served as a preliminary test case for the \geoclaw software in modeling the Teton Dam failure.

\subsubsection{Malpasset Dam failure}

\paragraph{\geoclaw study}
The \geoclaw study of Malpasset Dam failure by George was the first to apply \geoclaw to dam failure modeling \cite{george2011adaptive}. The preliminary results for the first validation of this code for dam-break flooding problems demonstrated that \geoclaw is a viable alternative to using specially developed unstructured meshes exhibiting minimal computational cost with result accuracy because of AMR’s efficiency \cite{george2011adaptive}. This study compared \geoclaw results to results of the commercial software package TELEMAC-2D \cite{valiani2002case, hervouet1999malpasset}. The comparison concluded that all three numerical simulations had nearly identical results that differed more from laboratory data than from one another \cite{george2011adaptive}.

\paragraph{\hecras study}
\hecras v4.1.0 was used to model the Malpasset dam breach benchmark problem compared to 2D modeling software Flood Modeller (formally known as ISIS) \cite{almassri2011numerical}. In simulating the dam break test, \hecras produced results similar to the historical numerical values.  The study concluded that \hecras was efficient, fast, and accurate for simulating dam breaches. However, both Flood Modeller and \hecras models required a sizable amount of data input to initiate the models and produce reliable results \cite{almassri2011numerical}. 

\section{Methods and Data} 

\subsection{Metadata and Domain}

This study parameterizes the dam geometry, reservoir, and domain as inputs for the numerical simulations. We use a large domain, the Teton Dam study area (88.5 km x 43.5 km), to evaluate the performance of a uniform \manningsn (surface roughness coefficient) and the software's handling of the complex terrain-the steep-walled narrow canyon and the floodplain.  This study selected a downstream limit of the model at \SI{54}{\km}, or just past the town of Roberts, Idaho (US).  This study chose the downstream limit based on the limitation of high-resolution topographic data. However, this limitation does not prevent us from a complete study comparing the \geoclaw and \hecras models.  In \hecras we also parameterize the earthen dam piping failure to appropriately simulate the time-dependent failure sensitivity analysis. 
  
For this benchmark problem parameterization, horizontal and vertical datum conversions were used to align topographic features with historical values. Two topographies were implemented in this modeling effort, originally sourced from the USGS in projection WGS 84 \cite{USGS_2015_repo}: "TetonLarge" (6.99 KB) at \SI{30}{\meter} resolution and "TetonHighRes" (7.77 KB) at \SI{10}{\meter} resolution. TetonHighRes and TetonLarge form a \SI{88.51} x \SI{43.45}{\cubic\kilo\meter} domain. The joined topography was used in the \geoclaw and \hecras models to enable model comparison.

\subsection{Reservoir volume}
This study estimated the initial reservoir volume using historical design drawings and updated volume estimates to approximate the reservoir extent \cite{chadwick1976report, randle2000geomorphology}.  Three reports by Reclamation provide three different estimates of the reservoir volumes at time of failure: \SI{3.08e8}{\cubic\meter} - \citeyear{chadwick1976report}, \SI{2.89e8}{\cubic\meter} - \citeyear{randle2000geomorphology}, and \SI{3.10e8}{\cubic\meter} - \citeyear{reclamation2008}. The values seem similar but may be significant in downstream inundation depths; this study tests the model's sensitivity to the reservoir volume. Each of these calculations was conducted differently. Balloffett and Scheffler \citeyear{balloffet1982numerical} suggest that the \citeyear{chadwick1976report} Reclamation's initial reservoir volume was calculated using an elevation-storage curve, implying a horizontal reservoir surface which additional work determined to not significantly affect 1D flood routing \cite{thomas1977flood, fread1977flood, chadwick1976report}. In comparison, Reclamation improved volume estimates in \citeyear{randle2000geomorphology} by using higher resolution contour maps to calculate volume from contour elevation maps. 

These volume estimates were compared with calculations from Google Earth using the historical parameters for the reservoir to map the quasi-prismatic river channel in sections. Spero et al. calculated the average height, and obtained an upper bound estimate of \SI{3.51e8}{\cubic\meter} \cite{speromodeling}. The volume discrepancies show the necessity for this study to determine the influence of reservoir volume on downstream flow using the 2D \hecras software and sensitivity analysis. The Teton Dam reservoir elevation was parameterized at an initial elevation of \SI{1617}{\meter}.

\subsection{\geoclaw Simulation}
The first step in constructing the \geoclaw simulation involved determining key input parameters and creating the reservoir. For the \geoclaw model, this study uses values from \citeauthor{randle2000geomorphology} because of their datum and more accurate volume calculation method \citeyear{randle2000geomorphology}. To parameterize the reservoir, a polygon was defined at Teton Canyon's edges, extending \SI{16}{\kilo\meter} upstream from the dam site. The \geoclaw ‘reservoir’ polygon was used to estimate the volume and flow. Points were plotted at the historic dam height throughout the reservoir – \SI{1624}{\meter} MSL. Then, the Teton River was filled with water to \SI{1617}{\meter} water surface elevation (NAVD 88) \cite{randle2000geomorphology}. The reservoir volume in \geoclaw was \SI{3.05e8}{\cubic\meter}, considered to be within suitable range.

\geoclaw was parameterized using an instantaneous dam breach assumption; the first time step of the simulation represents the dam's immediate removal. We also enforced refinement by associating minimum flow criteria with maximum mesh resolution, ensuring the reservoir was refined enough to provide reasonable resolution to capture the moving flood front. The Manning's coefficient for the study area was set to be a constant of 0.06. This value was estimated from a field inspection of the floodplain and Teton River Canyon. For confirmation of the \manningsn value, we used \manningsn definitions \cite{chow1959open}, gauge data, high watermark data, field interviews, newspaper records, high water marks, and hydrograph comparisons, and verification data to determine the confidence of 0.06 value \cite{gundlach1977guidelines}.  

Both stationary and Lagrangian gauges were inserted into the \geoclaw simulation, tracking flood wave arrival time and depth downstream.  In \Tab{gauges}, gauge locations are provided.  Stationary gauges output the SWE solution at fixed geographical positions and serve as proxies for comparison with historical records and accounts and log the flood wave arrival times and peak flow values. 

To visualize the flow field, we create a series of output images (.png files), viewed in the Google Earth browser. The color scale on the images is used to depict inundation depth.

\begin{table}
\begin{center}
\caption{Stationary Gauge Locations in \geoclaw Teton Dam domain.}
\begin{tabular}{cccc}
    \toprule
    Gauge Name & Latitude & Longitude \\
    \midrule
    Teton Canyon & -111.5939 & 43.9341  \\
    Teton Canyon Mouth & -111.6664 & 43.9338  \\
    Wilford & -111.6721 & 43.9144 \\
    Sugar City &-111.7601 & 43.8633 \\
    Rexburg  & -111.7923 & 43.8231 \\
    \bottomrule
\end{tabular}
\end{center}
\label{tab:gauges}
\end{table}

Tracking the fluid continuum downstream was difficult around localized topographic highs (ex: Menan Butte \SI{1713}{\meter}), so this study used two $3 \times 3$ grids of Lagrangian particles \ignore{(Supplementary Files; \Tab{software})}. \ignore{Our study is the first to use Lagrangian gauges in \geoclaw for terrestrial flow modeling.} The trajectories of the Lagrangian particles demonstrate the fixed velocity at points in space. The passively advected tracer particles move according to interpolated Eulerian velocity fields. We visualize the particles on top of water particles through (1) the Fundamental Principle of Kinematics (the velocity at a given position and time is equal to the velocity of the parcel that occupies that position at that time), (2) The material or substantial derivative relates the time rate of change observed following a moving parcel to the time rate of change observed at a fixed position; where the advective rate of change is in field coordinates. 3) To assert the conservation laws for volume and momentum within an Eulerian system, we need to transform the time derivative of an integral over a moving fluid volume into field coordinates; this leads to or requires the Reynolds Transport Theorem \cite{price2004res}. More information for the \geoclaw particles is available at \cite{lagrangiangauges}.

\subsection{\hecras Simulation}
The first step to parameterizing the historic Teton Reservoir in \hecras is to develop a terrain data set in the \hecras RAS Mapper. The projection was set to the new raster projection for unit agreement using ArcGIS Pro 10.8.1 (US customary units were used within the \hecras model). The projection file was created with ArcGIS Pro 10.8.1. A new terrain was data set was created by layering TetonHighRes (\SI{10}{\meter}) on TetonLarge (\SI{30}{\meter}).  TetonHighRes topography has a finer resolution, so it was given a higher priority in the combined Terrain Layer. To develop the 2D computational mesh, a polygon boundary was drawn for the 2D Teton Reservoir area. The Teton Reservoir (denoted TDRES2D) nominal grid resolution of \SI{61} times \SI{61}{\meter} cells was used to build the \hecras computational mesh. 

Similarly, the 2D downstream flow area was constructed using the RAS Mapper geometry editor. 
The 2D downstream flow area was refined with each run to reflect the computational area where flow occurred, improving computational efficiency. The final 2D downstream flow area for both the instantaneous \hecras and the time-dependent \hecras models used a uniform Manning’s coefficient of 0.06. The sensitivity analysis for this study focuses on the importance of Manning’s coefficient. The base of the downstream flow area (2D-DSTREAM) fed into a boundary condition line (BC Line) where the water could exit the simulation. 
Additionally, two break-lines were inserted into the simulation to force a cell edge at two important lateral features, differences in elevation. Through break-lines, we are better able to simulate water flow over cells. The first break-line was within the Teton Dam canyon downstream of the Teton Dam. Break-line 2 was inserted approximately from (\SI{43.98},\SI{-111.58}) to (\SI{43.90}, \SI{-111.65}). 

The two 2D Flow Area elements (TD-RES-2D and TD-DSTREAM) are connected with a storage area 2D connector (SA/2D), which formed the dam as a weir or embankment. The dam was parameterized in the geometry editor for the instantaneous dam failure by modeling the weir to fit the terrain. The time-dependent dam failure was also parameterized in the geometry editor. The dam structure was built using the historic dam height of \SI{1626}{\meter} (NAVD 88) \cite{chadwick1976report}. 

\begin{figure}
    \centering
    \includegraphics[width=\textwidth]{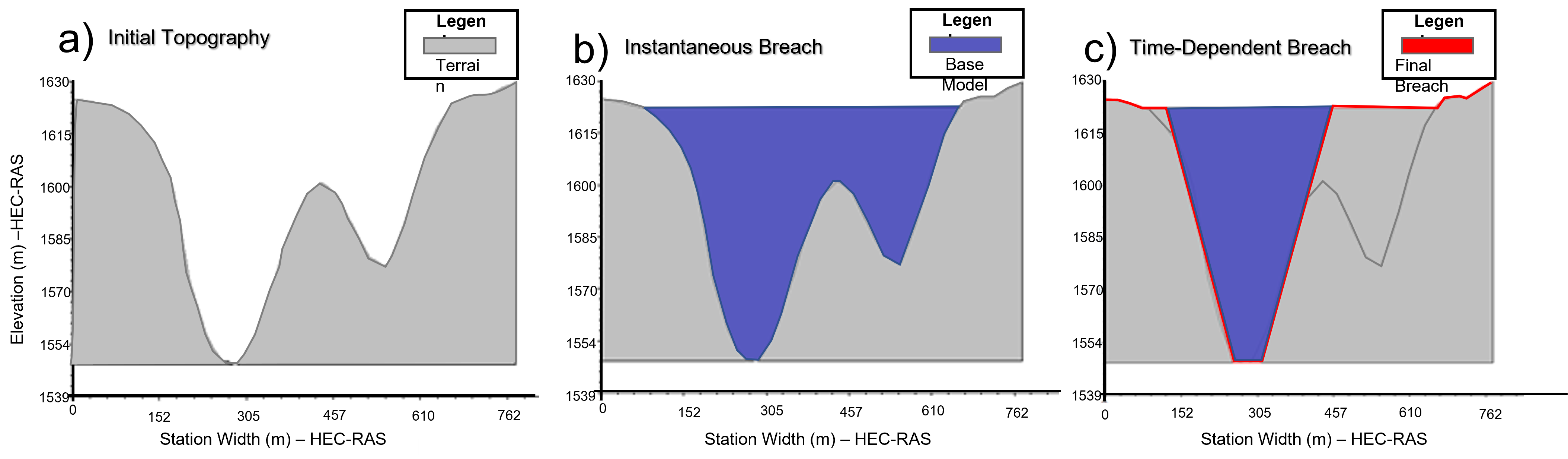}
    \caption{HEC-RAS model of parameterizing the dam breach.\textbf{a)} Initial topography with the terrain illustrated in gray, resolution of 10 m from TetonDamLatLong.topo topography. \textbf{b)} The base model is denoted in blue. The left area is the eroded historical failure area. Whereas, the right area denotes the location that Reclamation removed post-disaster \cite{magleby1981post}. \textbf{c)} The time-dependent final breach polygon is outlined in red. Historical values were implemented to match historical literature values and fit the terrain using the Breach (plan data) section of the SA/2D connector. The best fit was identified as having a Center Station of \SI{284}{\meter}, a final bottom width of \SI{55}{\meter}, and a final bottom elevation of \SI{1549}{\meter} (NAVD 88).}
    \label{fig:breach}
\end{figure}

For additional time-dependent dam failure parameterization, the piping coefficient was 0.5 based on \cite{chadwick1976report}, and the initial piping elevation from historical literature (first spot of seepage) was \SI{1597}{\meter} \cite{chadwick1976report}. The start date and time of the failure to achieve a maximum breach at 11:57 were and a simulation start time of 11:00; start time was chosen based on time stamps on photographs from 1976 failure. A sine wave breach progression was used as the breach progression method, validated in other literature studies \cite{bhandari2017one}. The progression method depicts how the breach grows from initiation to maximum size during the breach period (bp), where for this study, the bp is 11:00-12:00, one hour. The sine wave progression speed varies over the development time according to the first quarter cycle of a sine wave. It was chosen as it most closely resembles an earthen dam breach \cite{ackerman2008hydrologic, yochum2008case}.

\ignore{Historical values were implemented to match historical literature values and fit the terrain using the Breach (plan data) section of the SA/2D connector. The best fit was identified as having a Center Station of \SI{284}{\meter}, a final bottom width of \SI{55}{\meter}, and a final bottom elevation of \SI{1549}{\meter} (NAVD 88). For additional time-dependent dam failure parameterization, the piping coefficient was 0.5 based on \citeA{chadwick1976report}, and the initial piping elevation from historical literature (first spot of seepage) was \SI{1597}{\meter} \cite{chadwick1976report}. }

\ignore{The start date and time of the failure to achieve a maximum breach at 11:57 were and a simulation start time of 11:00; start time was chosen based on time stamps on photographs from 1976 failure. A sine wave breach progression was used as the breach progression method, validated in other literature studies \cite{bhandari2017one}. The progression method depicts how the breach grows from initiation to maximum size during the breach period (bp), where for this study, the bp is 11:00-12:00, one hour. The sine wave progression speed varies over the development time according to the first quarter cycle of a sine wave. It was chosen as it most closely resembles an earthen dam breach \cite{ackerman2008hydrologic, yochum2008case}.}

For the instantaneous dam failure plan, the simulation time starts at the time of breach, 11:57 on 5 June 1976 \cite{chadwick1976report} and runs for simulated time 12 hours and 43 minutes (24:00 6 June 1976). The computational interval and time stepping scheme depends on the Courant–Friedrichs–Lewy condition (CFL). Besides the geometry files, the primary difference in the two analysis runs was the simulation start times. The time-dependent dam failure required the sine wave initiation and the breach lasted 1 hour, the simulation begins before the dam breach (11:00) and runs until 24:00 6 June, 1976. Additionally, the first \hecras model was constructed using the Diffusion Wave equations. Then, once the model was stable, this study changed the model to solve using the Full Momentum SWE equations.

\section{Results of \geoclaw and \hecras Models}
Results are presented in three sections.  \Sec{results_a} focuses on stationary gauge results beginning upstream at the Teton Dam site and moving downstream until Rexburg (\SI{25}{\kilo\meter} downstream; \Tab{gauges}). \Sec{results_b} focuses on the lateral extent of flooding. \Sec{results_c} focuses on Lagrangian gauge components in \geoclaw and results. 

\subsection{\geoclaw and \hecras Instantaneous Dam Failure Results}
\label{sec:results_a}

\ignore{The ability to quantify, forecast, and calculate the downstream consequences of dam failure is crucial for protecting communities downstream of dams.} We compare all historic inundation depths (maximum) against numerical simulations performed using our \geoclaw numerical dam model.  The results focus on five of the \geoclaw and \hecras stationary gauges, which logged historic arrival times (hrs) and maximum flow depth (m) of the flood during the model’s simulation time. The gauge results are presented below in sequential order moving from the dam progressively downstream: (i) Teton Dam Canyon gauge, (ii) Teton Dam Canyon Mouth gauge, (iii) Wilford gauge, (iv) Sugar City gauge, and the (v) Rexburg gauge.  The sensitivity analysis use results from three profile lines that logged flow and flood wave arrival times: (a) Sugar City, (b) Rexburg, and (c) Menan Butte Butte \Tab{profile}. 

\subsubsection{Teton Dam Canyon Gauge}

 The \geoclaw Teton Dam Canyon gauge showed a \SI{7.6}{\meter} maximum depth flood wave, averaging to \SI{7.3}{\meter}, propagating down the canyon. Because \geoclaw models an instantaneous dam breach, the inundation flood wave arrival time for the \geoclaw Teton Dam Canyon gauge occurs during the second time step – almost immediately at 12:05. In comparison, photographs at the time of failure at the location of the \geoclaw Teton Dam Canyon gauge demonstrate a maximum flood wave depth of about \SI{15}{\meter} and arrival times at 12:05 \cite{chadwick1976report}. The \hecras Teton Dam Canyon Gauge shows \SI{41.3}{\meter} maximum depth flood wave, within the Teton Canyon. The flood wave arrives at the \hecras gauge at 11:59, just two minutes following the instantaneous dam breach at 11:57 \cite{chadwick1976report}.
 
\subsubsection{Teton Dam Canyon Mouth Gauge}
 
The \geoclaw Teton Dam Canyon Mouth gauge records flood wave arrival time five minutes later at 12:10. The \geoclaw gauge registers a maximum flood wave depth of \SI{2.1}{\meter} inundation, which is considerably less than the historical values of \SI{11.8}{\meter}-\SI{12.2}{\meter} \cite{chadwick1976report}. Then, as the flood laterally spread out of the canyon, it does not flood the town of Teton, agreeing with the lateral extent of the historic flood. However, the historic depths are greater than those modeled in \geoclaw by \SI{9.2}{\meter}. For the \hecras Teton Dam Canyon Mouth Gauge, the maximum recorded depth was \SI{20.6}{\meter}, \SI{8.5}{\meter} greater than historical values \cite{chadwick1976report}. The \hecras gauge logged a flood wave arrival time of 12:06.

\subsubsection{Wilford Gauge}

At 12:38, the \geoclaw Wilford gauge displays a depth of \SI{2.7}{\meter}. Historical literature values show flood waves reaching Wilford at approximately 12:45 with \SI{3.1}-\SI{4.6}{\meter} inundation depth \cite{chadwick1976report}. The \hecras model Wilford gauge displays maximum inundation depth of 
\SI{7.2}{\meter} overestimating the \SI{3.1}-\SI{4.6}{\meter} in historical data by about \SI{2.6}{meter}. Additionally, \hecras had an arrival time at 12:34. 

\subsubsection{Sugar City Gauge}

The \geoclaw Sugar City gauge showed a flood wave arrival time at 13:05. The historical literature value for arrival time was 13:30, and flood depth was \SI{3.1}{\meter} \cite{chadwick1976report}. The depth, as displayed in the \geoclaw gauge, is \SI{1.2}{\meter}. The \hecras Sugar City gauge registered a flood wave arrival time at 14:05 \cite{chadwick1976report} and a maximum depth of \SI{3.4}{\meter}. 

\subsubsection{Rexburg Gauge}

The \geoclaw Rexburg gauge demonstrated model values between \SI{1.2}-\SI{1.5}{\meter}; historical depths were \SI{2.4}{\meter}\cite{chadwick1976report}. The Rexburg \geoclaw gauge logs an arrival time at 14:30, and the historic arrival time is 14:30 \cite{chadwick1976report}. The \hecras Rexburg gauge shows an arrival time of 16:25, an hour and 55 minute difference from historical values. The gauge also logs a maximum depth of \SI{34}{\meter}.

\subsubsection{Lateral Extent of Modeled Floods and Computational Costs}

The other evaluation principle for the comparison criterion for \geoclaw was determining the lateral flood extent. The \geoclaw model showcased a flood area covered \SI{313}{\square\kilo\meter} which is within ± \SI{80}{\square\kilo\meter} of the historic inundation extent of \SI{337}{\square\kilo\meter} for the time simulated \cite{chadwick1976report}. In comparison, the \hecras model demonstrated a flood area of \SI{451}{\square\kilo\meter}, which is also within \SI{80}{\square\kilo\meter} of the historical \SI{337}{\square\kilo\meter} \cite{chadwick1976report}.

\geoclaw computational time for runs on the R2 compute cluster installed at Boise State University, is 17 minutes processing time and 15 minutes for plotting. 
For \hecras the base model run time was 31 minutes, which includes both run and plot time. 
Therefore, \geoclaw and \hecras have similar computational wall clock times, 32 minutes compared with 31 minutes. 
The \geoclaw model ran on 24 OpenMP threads on a single Intel Xeon Gold 6252 processor (2.10 GHz) of the R2 super compute cluster {\cite{bsurc2017r2}}. The computational budget of the \hecras Teton Dam base model included 28 cores on a single Intel Xeon E5-2680 v4 14 core (2.4GHz) node.  \hecras simulations were run on an Intel Xeon E5-2680 v4 14 core 2.4GHz (x2).

\begin{figure}
\centering
\includegraphics[width=14cm]{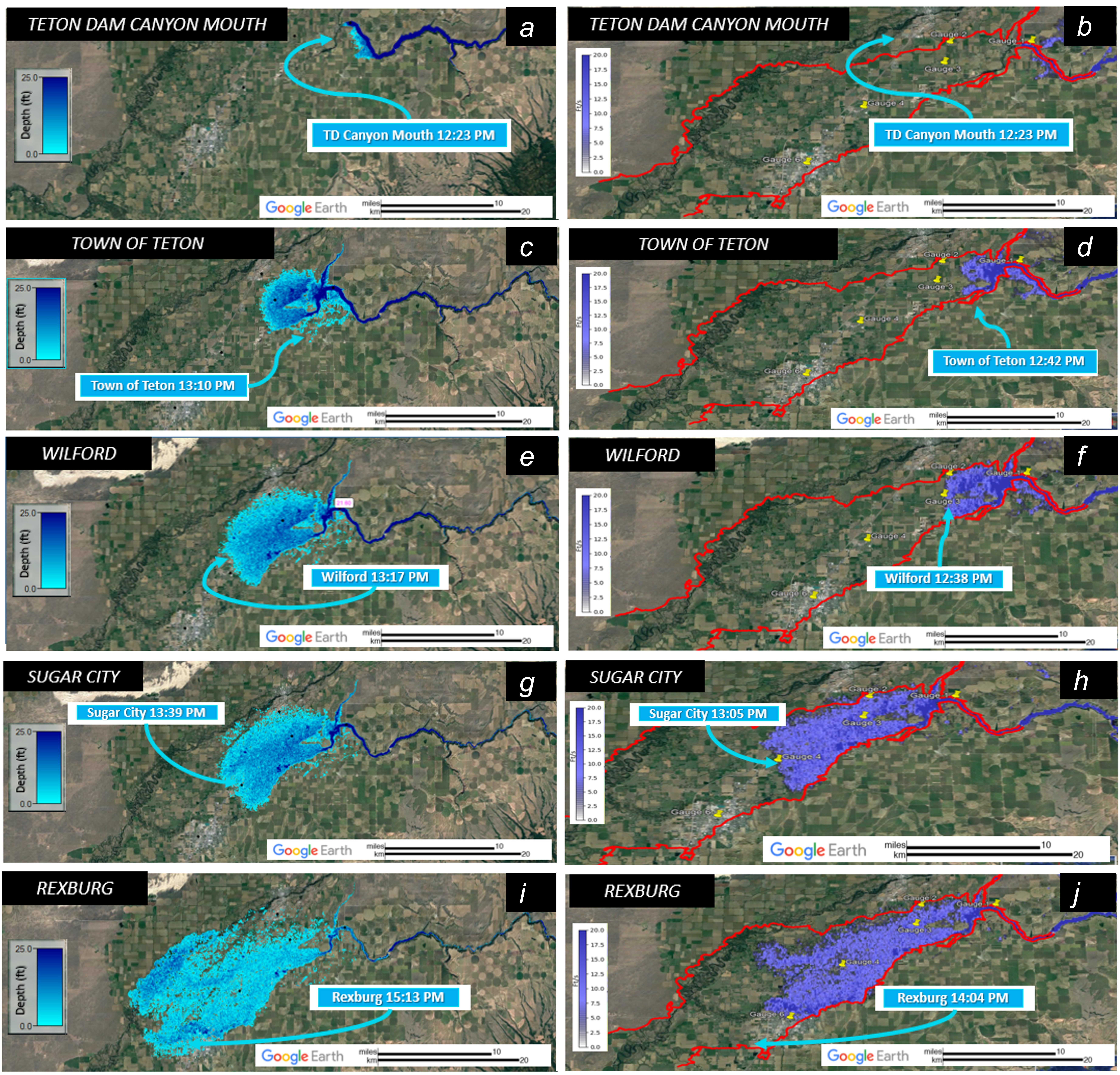}
\caption{\geoclaw results compared with historical data and \hecras results; focused on lateral extent of the flood (blue). The red outline denotes the extent of the historical flood from \cite{usgs1976survey}, and the yellow pushpins in the \geoclaw model results show the specific gauge locations; latitudes and longitudes of the gauges can be found in \Tab{gauges}. Figures a-e move progressively downstream from the dam, and a,c-e are gauges used in model comparison.}
\label{fig:geoHECresults}
\end{figure}

\subsubsection{Lagrangian Gauges Results}
\label{sec:results_c}

\begin{figure}
\begin{center}
\includegraphics[width=0.5\textwidth]{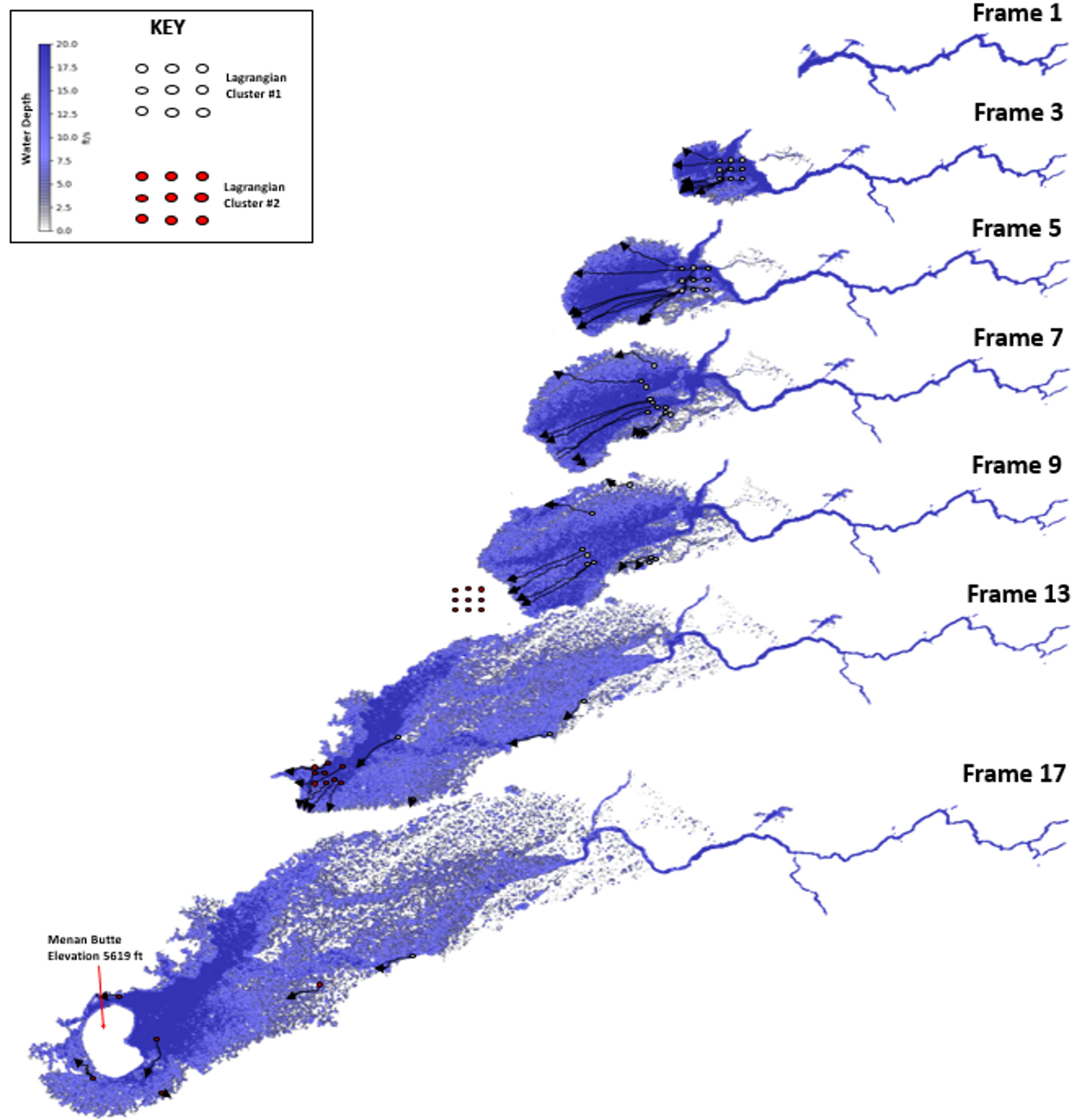}
\end{center}
\caption{GeoClaw flow progression from 11:57 5 June 1976 (Frame 1), until 00:00 6 June 1976 (Frame 17). Two Lagrangian clusters, Teton Dam Canyon Cluster and Menan Butte Cluster,  and their corresponding flow paths are denoted (white and red respectively, flow paths are in black)). Chaotic and turbulent eddying is observed around Menan Butte, ID (Frame 17)}. 
\label{fig:lagrangian}
\end{figure}

This study introduced Lagrangian particles into our depth-averaged flow field in the \geoclaw runs to better image the downstream flows for indicating turbulence.  Results, shown in \Fig{lagrangian} indicate swirling flow dynamics were observed in the dam failure simulation behind the local topographic high of the domain, Menan Butte (elevation \SI{1713} MSL).  As the Lagrangian gauges updated at each time step, both clusters demonstrated interesting flow paths within the downstream flood. For example, eddying occurred upstream of Menan Butte and detained six of nine Lagrangian gauges for over 15 minutes. Menan Butte is a large topographic feature, and thus results showed the flow was substantially affected below the Henry's Fork river, where the Snake River begins in Eastern Idaho.  At Henry's Fork, this study noted increased velocities for particles that interacted with the river, as they moved further downstream per time-step than other particles from the same cluster moving over the farmland domain.

\subsection{\hecras Sensitivity Analysis}
\label{sec:results_b}

This study used four sensitivity analyses to evaluate which parameters in the \hecras model control the numerical solutions (Table 3). For the \hecras model results, we analyzed the output flow hydrograph head and tailwaters, the computation log for error percentage, the gauge features for water surface elevation, and three profile lines for flow.  The three profile lines, Sugar City, Rexburg, and Menan Butte were chosen based on relative distance downstream and the plethora of historical data. \Tab{profile} summarizes the four sensitivity analyses:
\begin{enumerate}
    \item Manning’s roughness coefficient
    \item Volume Analysis
    \item Instantaneous and Time-Dependent Dam Failure
    \item Characteristic Size of the Computational Mesh
\end{enumerate}

Critical trends in the results demonstrate that the reservoir volume controls the peak flow but not the peak flow arrival time.  In contrast, the computational mesh controlled peak flow arrival time but had similar peak flows. 
The Manning’s $n$ value was likely overestimated in this study at 0.06, exhibited in the Manning’s sensitivity study. Lastly, the instantaneous dam breach assumption was validated as the base model values were identical in peak flow arrival time and comparable in peak flow. 



\begin{table}
\begin{center}
\caption{\hecras Sensitivity Analyses including the (i) \manningsn models, the (ii) reservoir volume models, the (iii) Characteristic Mesh Size models, and the (iv) instantaneous and time-dependent dam breach models. Table includes historic values to reference, sourced from \cite{usgs1976survey, chadwick1976report, usgshistoricalgage}.}
\begin{tabular}{cccccccc}
\toprule
\multicolumn{8}{c}{\hecras Sensitivity Analysis} \\
\midrule
& & \multicolumn{2}{c}{Sugar City} & \multicolumn{2}{c}{Rexburg} & \multicolumn{2}{c}{Menan Butte} \\
\cmidrule(lr){3-4}\cmidrule(lr){5-6}\cmidrule(lr){7-8}
& & {Flow} & {Arrival} & {Flow} & {Arrival} & {Flow} & {Arrival}\\
\midrule
\multicolumn{2}{c}{Historical Values}
& {\num{30016}} & {13:30} & {---} & {14:30} & \num{2265} & {---}\\
\midrule
\multirow{5}{2.1cm}{\manningsn} 
& 0.03 & \num{35458} & 13:18   & \num{21836} & 14:41   & \num{7850} & 16:07 \\
& 0.04 & \num{30957} & 13:34   & \num{18517} & 15:16   & \num{6391} & 17:05 \\
& 0.05 & \num{27366} & 13:50   & \num{16005} & 15:51   & \num{5363} & 18:03 \\
& 0.06 & \num{24488} & 14:05   & \num{14046} & 16:25   & \num{4611} & 19:01 \\
& 0.07 & \num{27745} & 14:20   & \num{12485} & 17:00   & \num{4036} & 19:39 \\ 
\midrule
\multirow{2}{2cm}{Reservoir Volume} 
& \num{2.29e8} & \num{24488} & {14:05} & \num{14046} & {16:25} & \num{4611} & {19:01} \\
& \num{2.63e8} & \num{24488} & {14:05} & \num{14046} & {16:25} & \num{4611} & {19:01} \\
\midrule
\multirow{3}{2cm}{Mesh Cell Size}
& {Base Model}  & \num{24488} & {14:05} & \num{14046} & {16:25} & \num{4611} & {19:01} \\
& {50\% Incr.}  & \num{22979} & {14:14} & \num{11111} & {17:04} & \num{3962} & {20:20} \\
& {50\% Decr.}  & \num{22968} & {14:14} & \num{11106} & {17:04} & \num{3963} & {20:20} \\
\midrule
\multirow{2}{2cm}{Dam Failure Mode}
& Instant.         & \num{24488} & {14:05} & \num{14046} & {16:25} & \num{4611} & {19:01} \\
& Time-Dep.        & \num{24488} & {14:05} & \num{14046} & {16:25} & \num{4611} & {19:01} \\   
\bottomrule
\end{tabular}
\end{center}
\label{tab:profile}
\end{table}

\subsubsection{Manning's Roughness Coefficient Sensitivity Analysis}

The Manning’s $n$ analysis compared roughness values of 0.03-0.07 to determine if the Manning’s $n$ value controlled the numerical solution.  The base model used a value of 0.06, and all Manning’s $n$ values had a slightly different mesh than other models with an expanded domain laterally to allow for natural flow instead of mesh-directed flow. At the Sugar City profile line, the 0.03 Manning’s model peak flow arrived first at 13:18, followed sequentially by progressively smaller Manning’s (0.04, 0.05, 0.06, and 0.07).  The peak flow was largest at the Manning’s 0.03 model (\SI{35300}{\cubic\meter/s}) and smallest for the Manning’s 0.07 model (\SI{22000}{\cubic\meter/s}).  At the Rexburg profile line, these trends continued as the Manning’s 0.03 model peak flow arrived first (14:41), and the Manning’s 0.07 model peak flow arrived last (16:25) – a difference of 1 hour and 45 minutes. Historically, the wave arrived in Rexburg at 14:40 \cite{chadwick1976report}.  The peak flow at the Rexburg profile line was largest for the Manning’s 0.03 model (\SI{21800}{\cubic\meter/s}) and lowest for the Manning’s 0.07 model (\SI{12500}{\cubic\meter/s}). The 0.06 \manningsn model (this study's base model) arrived at 16:25 with a flow of \SI{14,000}{\cubic\meter/s}.  Further downstream, at the Menan Butte profile line, the Manning’s 0.03 model peak flow arrives 16:07, and the Manning’s 0.07 model peak flow arrives 19:01. The Manning’s 0.03 model peak flow was observed as \SI{7900}{\cubic\meter/s}, whereas the Manning’s 0.07 was recorded as \SI{4000}{\cubic\meter/s}.  The base model, Manning's 0.06, recorded \SI{4600}{\cubic\meter/s} at 19:01.  The peak flow between the largest (0.07) and the smallest (0.03) \manningsn models differ by about 23\%. Results are shown in \Fig{mannings-n-sens}.

\ignore{
\begin{figure}
\centering
\includegraphics[width=14cm]{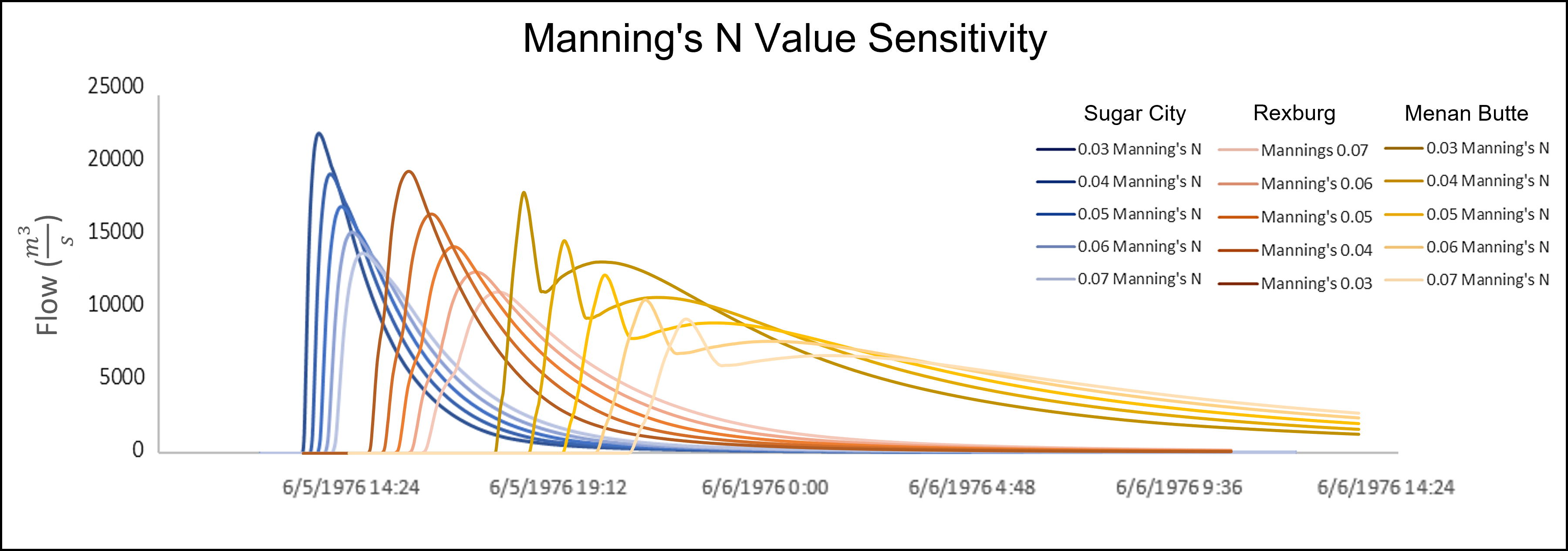}
\caption{\manningsn sensitivity analysis results from three profile lines in the \hecras domain: Sugar City, Rexburg, and Menan Butte. The Menan Butte flow hydrograph has irregular shape from expected bell curve as the flow encounters the topographic high (Menan Butte) and pools and eddies behind it.}
\label{fig:mannings-n-sens}
\end{figure}
}

\begin{figure}
    \centering
    \includegraphics[width=13cm]{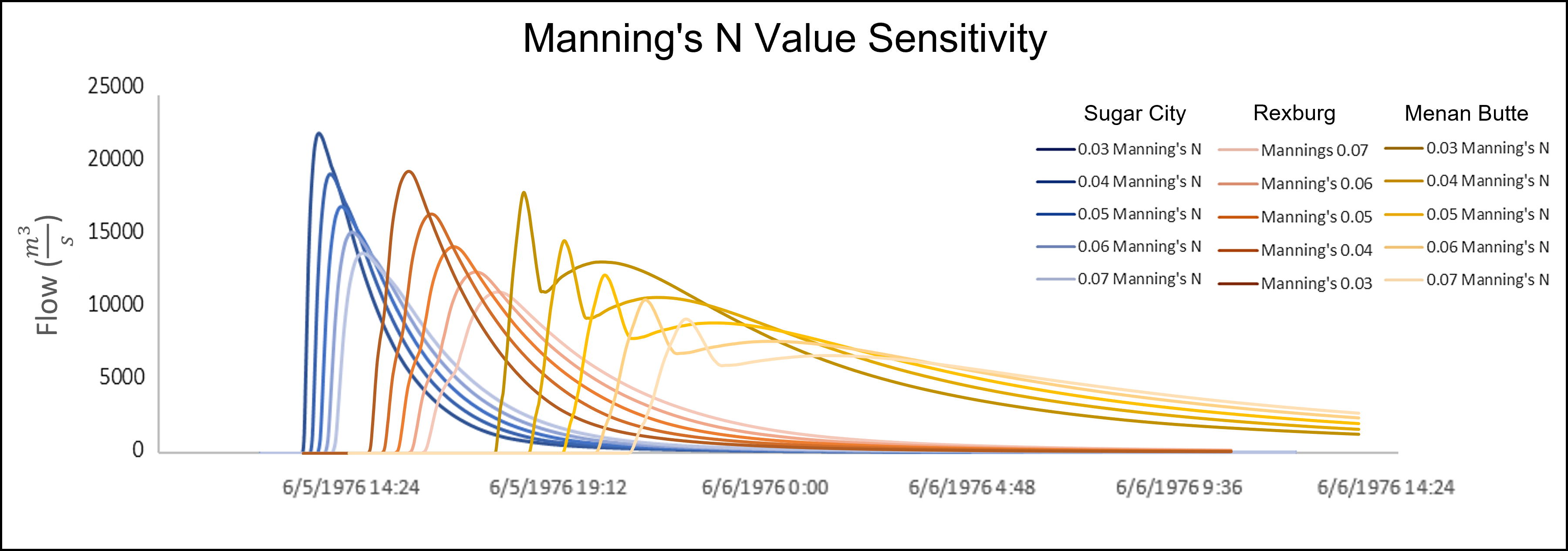}
    \caption{HEC-RAS base model sensitivity to Manning's roughness coefficient (0.03-0.07) observed at three profile lines in the domain (listed from upstream to downstream): Sugar City (blue), Rexburg (orange), and Menan Butte (yellow). }
    \label{fig:mannings-n-sens}
\end{figure}

\subsubsection{Volume Sensitivity Analysis}

The volume sensitivity analysis compared two intial reservoir volumes. Volume 1 is \SI{2.29e8}{\cubic\meter} (\SI{235978} acre-ft) representing the historical Reclamation volume value \cite{randle2000geomorphology}. Volume 2 created using the initial reservoir depth of  \SI{617}{\meter} water surface elevation (WSE; NAVD 88)  or \SI{2.63e8}{\cubic\meter} (\SI{213294} acre-ft model (\hecras Base Model). For the first profile line, Sugar City, the base model (\SI{617}{\cubic\meter}/s) arrived at 14:05, similar to the historic volume model (13:58). The peak flow at Sugar City between the two models differs by \SI{3200}{\cubic\meter/s}; historic volume model flow of \SI{27600}{\cubic\meter/s} and base model flow of \SI{24400}{\cubic\meter/s}. At the Rexburg profile line, the base model arrived at 16:25, and the historic volume model arrived at 16:13. The peak flow rates differed by \SI{2400}{\cubic\meter/s}, with the historic volume model logging the higher flow at \SI{16400}{\meter^3/s}. Then, at the third profile line, Menan Butte, the volume analysis demonstrated the historic volume model to arrive first at 18:38, and then the base model arrived at 19:01, a difference of fewer than 25 minutes. The peak flow rate for the models were comparable, with \SI{4600}{\cubic\meter/s} (base model) and \SI{5300}{\cubic\meter/s} (historic model), which differ by about \SI{674}{\cubic\meter/s}.

The peak flow differs between the historic volume model (Volume 1) and Volume 2 by 6\% at the Sugar City profile line, 8\% at the Rexburg profile line, and 7\% at the Menan Butte profile line. The flood wave arrival times differ by about 0.4\% at the Sugar City profile line, 0.6\% at the Rexburg profile line, and 1\% at the Menan Butte profile line.

\subsubsection{Characteristic Size of Computational Mesh Analysis}

The results of the computational mesh analysis compare the base model mesh (\SI{242210} cells; \SI{274} x \SI{274} m - base model mesh) to a mesh with cell sizes increased by 50\% (\SI{11157} cells; \SI{411} x \SI{411}{\meter} - lower resolution mesh [LRM]) and a mesh with cell sizes decreased by 50\% (\SI{980,030} cells; \SI{137}x\SI{137}{\meter}-higher resolution mesh [HRM]). The computational cost for the HRM was 4:20:15 hours (\SI{15615} seconds), in comparison to the LRM, which processed in 14:03 minutes (\SI{843} seconds).  The base model computes at 31:13 minutes (\SI{1873} seconds). The HRM runs 52\% slower than the base model cost, a significant increase in cost. 

For the profile line results, the LRM and HRM produced similar results. At the Sugar City profile line the LRM registered \SI{22858}{\cubic\meter/s} compared to the HRM \SI{22800}{\cubic\meter/s}; a difference of 0.1\% with both flows lower than the base model flow of \SI{24400}{\cubic\meter/s}, but only by 3\%.  At the Sugar City Profile line then, the base model arrives first at 14:05, followed by the LRM and HRM arriving both at 14:14. Both the HRM and LRM log a flow of \SI{11100}{\cubic\meter/s}, whereas the base model's flow was 11\% higher at \SI{14000}{\cubic\meter/s}. The arrival time of the HRM and LRM were both 17:04, and the base model was 16:25. Further downstream at the Menan Butte profile line, the LRM and HRM arrive again simultaneously (20:20) after the base model (19:01). The LRM and HRM flows were within \SI{1}{\cubic\meter/s} being \SI{4000}{\cubic\meter/s} respectively); 6\% below the base model flow estimate of \SI{4600}{\cubic\meter/s}.
  
\subsubsection{Instantaneous and Time-Dependent Dam Breach Model Results}

For instantaneous breach model versus time-dependent breach model sensitivity analysis, this study assesses the time-dependent scenario and the instantaneous dam breach scenarios in \hecras.  Processed flood wave arrival times accounted for the time-dependent breach start time of 10:00 differing by 1:57 hours from the instantaneous breach.  At Sugar City profile line, the instantaneous dam breach logs a flow of \SI{24400}{\cubic\meter/s} compared to the time-dependent dam breach value of \SI{243400}{\cubic\meter/s} a difference of 0.2\%; both arrive at 14:05. At the Rexburg profile line, both models arrive at 16:25 and log extremely similar values (time-dependent model: \SI{14000}{\cubic\meter/s}, instantaneous model: \SI{14000}{\cubic\meter/s}).  Then, at the Menan Butte profile line, both models arrive at 19:01 and log similar flow values. The time-dependent breach logged and the instantaneous breach both logged \SI{4600}{\meter/s} logged \SI{4600}{\cubic\meter/s}. Furthermore, both computational costs were similar varying only by 30 $\pm{ 2}$ minutes.

\section{Discussion}

Dams in the US are aging, and the frequency of dam failures increases with age. Floods are one of the most frequent and costly natural disasters the US faces – including dam failures flooding \cite{asdso_2021}. Dam failures pose a significant threat to human life downstream of the dam. One must first validate and benchmark the software using a historical dam failure assessment to forecast dam failures. One of the guiding questions of the study focused on evaluating \geoclaw for dam failure downstream modeling. The evaluation was based on mathematical formulation controlling flood movement, the capability to predict inundation extent and final flow depth through a numerical method, and the ease of use, performance characteristics, and tools for visualization and post-processing. The second guiding question investigated the model's sensitivity to its parameterization and uncertainty.

The stationary gauges used in the two models, \hecras and \geoclaw, allow comparison of flood wave depth and flood wave arrival time. This section discusses the first two gauges: Teton Dam Canyon gauge, Teton Dam Canyon Mouth gauge, and the Wilford gauge. The first gauge, Teton Dam Canyon, showed a flood wave arrival time at 12:05 as \SI{7.6}{\meter} depth. We consider the \geoclaw Teton Dam Canyon gauge to agree with historical records when evaluating the flood wave arrival time (12:05 historical arrival time) \cite{chadwick1976report}, but not when evaluating by depth as it underestimates the depth by \SI{7.62}{\meter}. The \hecras model overestimates the flow depth by \SI{5.18}{\meter} but agrees with the historical flood arrival time. The \geoclaw Teton Dam Canyon Mouth gauge underestimates the flood depth by \SI{9}{\meter}. 

In contrast, the \hecras model’s Teton Dam Canyon Mouth gauge overestimates the depth by about \SI{27}{\meter}, and the flood wave arrival time differs from historical records by over 30 minutes. These differences are significant, and the slow arrival time and overestimation of the depth suggest that Manning’s n value of 0.06 is too high in the canyon. However, it is important to note that the Teton Canyon presents the most challenging portion of the flow to model using the SWE. As within the canyon, the extreme vertical accelerations are not well captured by the SWE model.

Continuing the discussion of gauge results with the Wilford, Sugar City, and Rexburg gauges, this study finds \geoclaw and \hecras values similar to historical values.  At the Wilford gauge, the \geoclaw model estimated the depth within \SI{1.2}{\meter} of historical values and the flood arrival time within 15 minutes – which we consider to be excellent agreement. The \hecras model overestimates the flood depth by \SI{4.6}{\meter} and demonstrates a flood arrival time difference of 40 minutes which is significant (12:38 historical versus 13:17 model). The \geoclaw Sugar City gauge logged general agreement to the historical literature value arrival time (within 25 minutes) but underestimated the depth by 1.8 m. In contrast, the \hecras gauge estimated the flood within \SI{0.6}{\meter} of the historical value, and the flood arrival time is within 10 minutes – we consider the \hecras Sugar City gauge to be in excellent agreement with historical data. Further downstream, the \geoclaw Rexburg gauge also demonstrates inundation agreements with model values within \SI{0.9}{\meter} of historical values and the same model arrival time as historical arrival time (14:30). The \hecras Rexburg gauge slightly overestimated historical data; the arrival time differs by over 30 minutes.

Major trends in the results show that the \geoclaw model demonstrates good agreement with historical values for inundation extent, although consistently underestimating depth values. The base model results for \hecras showed agreement with both \geoclaw and with historical data, although consistently overestimating the maximum flow depth and arrival times. For lateral extent evaluation, the \geoclaw model was \SI{313}{\square\kilo\meter} which is within ± \SI{77.7}{\square\kilo\meter} of the historic inundation extent of \SI{336.7}{\square\kilo\meter}. However, the \hecras model overestimated the extent which could be related to overestimating the WSE in the reservoir; \SI{450}{\square\kilo\meter} \hecras base model compared to the \SI{337}{\square\kilo\meter} in historical archives, \cite{chadwick1976report}. 

In the assessment of the computational cost, both models are comparable with wall clock times between 15-17 minutes. \hecras post-processing requires no additional plotting, whereas \geoclaw requires about 15 minutes of extra processing to produce visualization output. Overall, the results allowed this study to answer the three guiding questions and come across ideas for future work. 

\subsubsection{Suitability of \geoclaw}

This study investigated if the \geoclaw model was suitable for dam failure downstream modeling of the Teton Dam failure (objectives outlined in Section 1). We determined the suitability of the \geoclaw software based on its capability to resolve lateral inundation extent, flood front arrival times, and maximum flood depths. For \geoclaw and \hecras, the calculated area for the lateral extent of the flood largely agrees with historical data. We evaluate the five inundated simulation domains and the modeled time of the six stationary gauges inserted into the \geoclaw simulation for gauge data interpretation.  We consider the model in excellent agreement if it predicts the flood wave arrival time within 15 minutes of historical data. We consider the model in good agreement if it predicts the wave within 30 minutes of historical data. Three of the five \geoclaw gauges log flood wave arrival times within ± 7 minutes (Teton Canyon, Teton Canyon Mouth, Wilford -excellent agreement), and all five gauges demonstrate good agreement with historical arrival times. By contrast, the \hecras base model shows three gauges in excellent agreement, one gauge in good agreement, and the Rexburg gauge predicting an arrival time two hours after the historic wave arrived (14:30 compared to 16:25). 

For assessing maximum flood depths, this study considered a model in excellent agreement if the depth values were within \SI{1.5}{\meter} of historical values. The \geoclaw model is in good agreement if values are within \SI{3}{\meter} of historical values. These values also take into account uncertainties in the model, such as ambiguity as to locations where historical values were collected \cite{chadwick1976report}. For \geoclaw, two gauges (Wilford and Rexburg) demonstrate excellent agreement with historical data, and Sugar City demonstrates good agreement with historical values. For \hecras, only Sugar City demonstrated excellent agreement, and Rexburg demonstrated good agreement. The Wilford gauge was overpredicted by \hecras (\SI{7.2}{\meter} compared to the historic \SI{4}{\meter}). The trends of \geoclaw show it consistently underestimates the maximum flood depths in the confined canyon area (by \SI{7.6} - \SI{9.1}{\meter}). However, as the modeled \geoclaw flood wave moves downstream out of the canyon, it largely agrees more with maximum flow depths, illustrating perhaps a resolution limitation of \geoclaw in the steep canyon terrain. Potentially, for both \hecras and \geoclaw, using a different roughness coefficient in the canyon could lead to improved and more realistic maximum flow depths.

Other factors considered in model results encompass flood volume, mesh refinement, code efficiency, and pre-processing and post-processing workflows. The pre-processing workflow for \geoclaw involved retrieving metadata in ASCII format and transferring that to the Boise State Compute Cluster R2 \cite{bsurc2017r2}. In contrast, \hecras topography processing required changing ASCII data to raster using ArcMap and ArcGIS Pro. When considering mesh refinement, \geoclaw’s AMR is advantageous, only resolving where the flood propagates and not requiring an iterative approach to refining the user-defined mesh to improve the downstream mesh. For code and run time efficiency, \hecras and \geoclaw have similar run times. With semi-parallel processing, \geoclaw could improve run times to similar processing and plotting speeds. For example, future work could use the ForestClaw package which could run \geoclaw using distributed computing \cite{calhoun2017forestclaw}. For the \hecras post-processing workflow, as the model runs, the RAS Mapper window simultaneously updates and stores all results allowing for visualization over the terrain or online map imagery.

\subsubsection{What is the importance of breach progression in dam failure modeling?}

The importance of a dam breach was assessed by comparing the \hecras time-dependent breach model to the instantaneous breach model. Both models yielded extremely similar results, and the same flood wave arrival times at all three profile lines. The sensitivity analysis required almost identical flows. Therefore, with the similar outputs between the two models, we consider using an instantaneous dam breach estimate rather than parameterizing the historical time-dependent dam breach to be valid. 

\subsubsection{What is the uncertainty associated with using \hecras for dam failure modeling?}

Another objective of this study was to determine the sensitivity of the \hecras model. The results from the sensitivity analyses indicate the volume and the Manning’s roughness coefficient were shown to largely control the solution and introduce uncertainty into the model. 

For the volume sensitivity analyses, this study compared the historic volume of \SI{2.29e8}{\cubic\meter} \cite{randle2000geomorphology} to the base model volume of \SI{2.63e8}{\cubic\meter}. The results indicated that the peak flow values between the historic and base models were progressively aligned as the flood wave moved downstream and laterally expanded. From Sugar City (flow difference of \SI{3200}{\cubic\meter/s} to Menan Butte (difference of only \SI{2400}{\cubic\meter/s}). However, the peak flow rates did show a range, so knowing an accurate reservoir volume is critical to building a reliable dam failure model. 

In this study, the \manningsn value also controlled the solution. For instance, the Sugar City profile line logs the 0.04 n flow arriving at 13:34, which is closer to the historical value of 13:30 than the 0.06 n base mode which arrived at 14:05 (difference of 35 minutes). Further downstream, the Rexburg profile line logs the 0.03 n flow arriving at 14:41, close to the historical value of 14:30. However, the base model (0.06 n) arrives at 16:25, nearly 2 hours later – a significant difference. We recommend that future work involves a depth-variable Manning’s roughness coefficient or a variable Manning’s roughness coefficient throughout the domain, given that the Manning’s sensitivity analysis showed that the roughness wholly affects the simulation and computational results. 
	
Although previous studies demonstrated that the mesh could control a dam failure flood solution, we conclude from the mesh sensitivity analysis that the mesh does not control or impact the solution of the Teton Dam flood wave arrival time or flow. With the low relief of the terrain (with Menan Butte and the Teton canyon as the only exceptions), and the fairly high resolution cell-gridding, the mesh does not change the computational result. However, the geometry is a crucial source of potential error in any hydraulic model with uncertainty. For example, having a higher resolution topography \SI{5}{\meter}) rather than \SI{10}{\meter} for the dam's reservoir could improve reservoir volume predictions to be closer to historical values. 

Additionally, there is an uncertainty associated with the historical depth value, as they were not directly associated with a location (longitude and latitude) when collected in 1976. Therefore, with that data limitation, there is uncertainty associated with the historical values \Tab{historicdata}. Future work could include a survey in Eastern Idaho of remaining structures to document known high watermarks to improve this data limitation. Then, this uncertainty could be eliminated by integrating those locations as stationary gauges. 

The \hecras hydrologic model a widely used dam failure modeling software in the US. \hecras in the past has been used to model the Teton Dam failure using the 1D unsteady flow routing (1D SWE equations) to route an inflowing flood hydrograph through a reservoir \cite{rec_1980_land}. In this project, we expand on previous research, employing \hecras v.5.0.7 2D unsteady flow routing capabilities (Full Momentum SWE) for comparison to \geoclaw. This study found both \hecras and \geoclaw to produce similar numerical solutions and resultant simulations with numerical gauges depicting maximum flow depths and flood wave arrival times that largely agreed with historical data. With this study and the results, we would recommend using \geoclaw for forecasting or hindcasting downstream flow behavior from dam breach simulations.  

\subsection{Recommendations for Future Work}
\subsubsection{Drone Photogrammetry Generated Topographies in Dam Failure Modeling}

As some dams are in remote locations where only coarse DEMs exist, drone photogrammetry could be a valuable tool for creating supplemental high-resolution DEMs. We recommend that future work investigate high-resolution topography usage in \geoclaw dam breach modeling, focusing on resolution and efficiency. Initial testing demonstrated that drone photogrammetry-generated topography could be uploaded into the \hecras’s RAS Mapper as a terrain in GeoTiff format. Through uploading the three raster data sets in this study, \hecras can further import them within a single layer, which can be merged into a single raster. Using \hecras or any ESRI product (ArcMap or ArcGIS Pro), exported combined terrains could be loaded into \geoclaw for a resolution and run-time efficiency-focused study. 

\subsubsection{Teton Dam \geoclaw Model Manning's Coefficient}
This study recommends additional Manning’s sensitivity analyses to be performed, comparing uniform \manningsn values which spatially varying Manning’s roughness sensitivity analyses and depth-averaged \manningsn. Additionally, to improve uncertainties in this model, we recommend quantifying the geomorphological differences in the canyon from 1976 pre-failure to the present day data as the volume of landslide debris that might be offsetting reservoir fill volume values in this study. Future work could involve simulation of a higher-resolution Teton Canyon (generated from drone photogrammetry as well) which might increase computation times, but could improve downstream canyon values.

\subsubsection{Depth-Averaged Debris Modeling of Teton Dam failure}

The current \geoclaw model uses the SWEs and Lagrangian particles to track streamlines. The massless Lagrangian particles could be further parameterized with mass, size, drag (function of particle Reynolds number), and buoyancy – the most significant forces acting on fluid objects. Through further parameterization, the user could model debris carried in the dam failure flood wave such as cattle, houses, sediment (sand), and timber. \geoclaw could use model development like COULWAVE and ComMIT/MOST(NOAA) to simulate buoyant debris. However, research explores the simulation of debris sourced from vegetation, vehicles, or non-buoyant debris (the additional parameterized Lagrangian gauges) such as buildings, sand, and rock. The Teton Dam failure provides an opportunity for future work in this area, which could help forecast dam failure risk and can assess costs as debris removal, improving community resilience.

\section*{Open Research}
The study used open-source software \hecras \cite{hecras2019release} and \geoclaw \cite{geoclaw58release} for 2D numerical modeling. The Teton Dam models (version used in this paper) are located on GitHub with a (i) READme.md file that includes the metadata and a complete model description, (ii) configuration parameters along with the specific script and workflow which is preserved in the repository, and the (iii) code which can produce the data that supports the summary results, tables and figures \cite{tdamrepository}. Additionally, this study used the high-performance computing support of the R2 compute cluster (DOI: 10.18122/B2S41H) provided by Boise State University’s Research Computing Department \cite{bsurc2017r2}. 

\subsection*{Data Availability Statement}

Both the (i) ASCII topography data used for creating the underlying terrain, and (ii) the \geoclaw and \hecras project files are available at [GitHub: Spero-Hannah/Teton-Dam-Failure-Example] via [DOI: 10.5281/zenodo.586668, persistent identifier link] with  [Berkeley Software Distribution (BSD) license]; \cite{tdamrepository}. Unmodified topography files can be found on the USGS website \cite{USGS_2015_repo}. 

\section*{CRediT Author Statement}
The authors confirm contribution to the paper as follows: 
Conceptualization: D. Calhoun; 
Methodology: D. Calhoun, H. Spero;
Software: D. Calhoun, H. Spero; 
Formal Analysis and Investigation: H. Spero; 
Resources: D. Calhoun, M. Schubert; 
Data Curation: D. Calhoun; 
Writing-Original Draft: H. Spero; 
Writing-Review and Editing: D. Calhoun, M. Schubert; 
Funding Acquisition: D. Calhoun, H. Spero.
All authors reviewed the results and approved the final version of the manuscript.

\acknowledgments
The Authors would like to thank Boise State University's Boise Center Aerospace Laboratory (BCAL) for access to software, the Idaho Technical Committee of the Northwest Regional Floodplain Managers Association (ITC-NORFMA), and others for their involvement and support of this research. The Authors would like to thank the Bureau of Reclamation and the Department of the Interior for allowing access and reproduction of their drone photogrammetry data and historical records. Further, the Authors would like to thank both Dr. Jim McNamara (thesis committee member) and the staff at HDR Engineering Inc. for their respective internal reviews of the manuscript. 
\textit{Land Acknowledgement} The Authors also acknowledge that the Teton River Canyon site area was home to the Shoshone and Bannock Indian people in the early 1800s. Per the “The Policy of the Shoshone-Bannock Tribes for Management of Snake River Basin Resources” the Shoshone-Bannock Tribes are acknowledged as the previous owners of the ceded lands of the Teton River Canyon study area \cite{reclamation2006}.

\textbf{Funding Acknowledgements}:
This work was supported by D. Calhoun's National Science Foundation [ \#1819257, 2018], “Parallel, adaptive Cartesian grid algorithms for natural hazards modeling"; H. Spero's Association of State Floodplain Managers student scholarship [2020]. Broader impacts of this research, a Virtual Reality Teton Dam environment, was awarded to H. Spero from Boise State University’s Undergraduate Research and Creative Activities Grant 2021 (URCA).




%


\end{document}